\pdfoutput=1
\documentclass{sigchi}

\toappear{Permission to make digital or hard copies of all or part of this work for personal or classroom use is granted without fee provided that copies are not made or distributed for profit or commercial advantage and that copies bear this notice and the full citation on the first page. Copyrights for components of this work owned by others than the author(s) must be honored. Abstracting with credit is permitted. To copy otherwise, or republish, to post on servers or to redistribute to lists, requires prior specific permission and/or a fee. Request permissions from Permissions@acm.org. \\
{\emph{CSCW '16}}, February 27--March 02, 2016, San Francisco, CA, USA. \\
Copyright is held by the owner/author(s). Publication rights licensed to ACM. \\
ACM 978-1-4503-3592-8/16/02 \$15.00. \\
DOI: http://dx.doi.org/10.1145/2818048.2820072}
\usepackage{balance}  % to better equalize the last page
\usepackage{graphics} % for EPS, load graphicx instead 
\usepackage{txfonts}
\usepackage{times}    % comment if you want LaTeX's default font
\usepackage[pdftex]{hyperref}
\usepackage{color}
\usepackage{textcomp}
\usepackage{array,booktabs}
\usepackage{ccicons}
\usepackage{tabularx}
\usepackage{verbatim}

\makeatletter
\def\url@leostyle{%
  \@ifundefined{selectfont}{\def\UrlFont{\sf}}{\def\UrlFont{\small\bf\ttfamily}}}
\makeatother
\def\pprw{8.5in}
\def\pprh{11in}

\definecolor{linkColor}{RGB}{6,125,233}
\hypersetup{%
  pdftitle={Storia: Summarizing Social Media Content based on Narrative Theory using Crowdsourcing},
  pdfauthor={Joy Kim, Andr\'{e}s Monroy-Hern\'{a}ndez},
  pdfkeywords={Social computing; crowdsourcing; creative collaboration; storytelling.},
  bookmarksnumbered,
  pdfstartview={FitH},
  colorlinks,
  citecolor=black,
  filecolor=black,
  linkcolor=black,
  urlcolor=linkColor,
  breaklinks=true,
}

\begin{document}

\title{Storia: Summarizing Social Media Content based on Narrative Theory using Crowdsourcing}

\numberofauthors{2}
\author{%
   \alignauthor{Joy Kim\\
     \affaddr{Stanford University}\\
     \email{jojo0808@stanford.edu}}\\
   \alignauthor{Andr\'{e}s Monroy-Hern\'{a}ndez\\
     \affaddr{Microsoft Research}\\
     \email{amh@microsoft.com}}\\
%  \alignauthor{Anonymized for submission\\
%    \affaddr{ }\\
%    \email{ }}\\
}

\maketitle

\begin{abstract}
People from all over the world use social media to share thoughts and opinions about events, and understanding what people say through these channels has been of increasing interest to researchers, journalists, and marketers alike.
However, while automatically generated summaries enable people to consume large amounts of data efficiently, they do not provide the context needed for a viewer to fully understand an event.
Narrative structure can provide templates for the order and manner in which this data is presented to create stories that are oriented around narrative elements rather than summaries made up of facts.
%Narrative structure can provide templates for the order and manner in which this data is presented to ensure essential storytelling elements---such as character introductions, a plot, and a resolution---are present in the story.
In this paper, we use narrative theory as a framework for identifying the links between social media content. To do this, we designed crowdsourcing tasks to generate summaries of events based on commonly used narrative templates. In a controlled study, for certain types of events, people were more emotionally engaged with stories created with narrative structure and were also more likely to recommend them to others compared to summaries created without narrative structure.
\end{abstract}



\section{Introduction}
Social media today allows millions of people from all over the world to share and discuss their thoughts about commonly experienced events. There has been increasing interest among researchers, journalists, and marketers alike in using social media to understand what people say about these events; a large body of research explores summarizing emotions and reactions as seen on Twitter \cite{Sharifi:2010:SMA:1857999.1858099,Nichols:2012:SSE:2166966.2166999,chakrabarti2011event}, and news articles and blog posts often integrate social media content and visualizations into their text.

However, while existing automated approaches excel at identifying moments of public attention, it is often up to the viewer to create their own interpretation of the data (or seek out interpretations provided by journalists and bloggers). This can be difficult because tweets, Facebook posts, and other social media messages are generally created in the moment \cite{Schirra:2014:TAM:2611205.2557070}, and so sometimes lack the context needed to make sense to future viewers. In addition, some viewers may not be familiar enough with the event to understand the jargon, idioms, or other specialized language used by social media authors.
%However, the sheer volume and fragmented nature of social media makes understanding content difficult; in response, automated techniques such as clustering content by topic for efficient browsing \cite{Sharifi:2010:SMA:1857999.1858099} and automatically generating text summaries of spikes in posting activity \cite{Nichols:2012:SSE:2166966.2166999,chakrabarti2011event} have been developed to highlight interesting moments and make them accessible without having forcing viewers to examine hundreds of individual posts. While these techniques address problems of accessibility and efficiency, they do not directly aid viewers in \emph{understanding} content. 

On the other hand, manually authored stories punctuated with a curated set of social media posts, such as those created through Storify \cite{storify}, can provide unifying commentary that bridge gaps in information: for example, a news article about Brazil's dramatic loss to Germany in the FIFA 2014 World Cup semi-finals describes not just the final score but also compares it against past matches to highlight the intensity of the loss; it also points out that Brazil's top scorer and top defender were unable to play during the match due to circumstances external to the match itself. However, the process for creating these stories is limited to those who have the time, skills, and resources to learn about an event's context by observing the event, doing research, and conducting interviews.

%tools, such as Storify \cite{storify}, help journalists and other skilled creators manually author stories that integrate social media posts. These stories fill holes left by purely automatic techniques by providing background, context, and imagery: a news article describing the 2014 Winter Olympics opening ceremony introduces the event not with an outline of performances to come but an interpretation of the show's message from Russia to the world; an article about Brazil's dramatic defeat to Germany in the 2014 World Cup semi-finals describes fans' shocked reactions, rather than just a list of goals. However, the process for creating these stories is often limited those who have the time, skills, and resources to learn about an event's context by observing the event or conducting interviews and write about the event in an engaging way. 

\begin{comment}
\begin{figure}[!t]
\centering
\includegraphics[width=1\columnwidth]{summary}
\caption{Storia guides crowdworkers in creating narrative summaries of social media events.}
\label{fig:summary}
\end{figure}
\end{comment}

% Insight + Approach
To achieve both the scalability of automated approaches and the coherency of manually authored stories, we propose \emph{summarizing social media content based on narrative structure}: rather than depending on a trained storytelling expert, interpreting social media in terms of narrative elements --- such as beginnings, middles, ends, characters, goals, and climaxes --- may reveal what information a summary needs in order to make sense to viewers. This approach makes use of crowdsourcing to interpret data at scale and automatically generate summaries with these narrative elements in mind.

We hypothesize that designing crowdsourcing tasks around narrative elements can help non-expert crowd workers collaborate in addressing this challenge.
%Through this, we extend past work --- machines are good at automatically detecting and generating structure in large datasets, and we use this lessen the amount of data crowd workers have to interpret and understand. 
In this paper, we first \emph{identify and fill narrative gaps} in a social media record. For example, if a name appears in a social media feed, we may want to know more about who they are and their significance with respect to the overall story (i.e. a ``character'' introduction). Then, we \emph{link content to narrative categories}---for example, we may be able to recognize certain tweets as descriptions of conflict between the story's characters. Specifically, we refer to \emph{narrative categories} \cite{cohn2013visual} as a simple template for structuring social media content with respect to storytelling roles. 

To explore this approach, we created a prototype crowdsourcing system called \emph{Storia} (Figure~\ref{fig:simple-data-driven-comics}). Storia takes, as input, data from an automated system that detects moments that occur during a public event (such as a sports game) and uses crowdsourcing to output a written story about the event.
Storia consists of two crowdsourcing modules that ask crowd workers to 1) gather missing narrative context and 2) write paragraphs for each important moment in the event based on social media content, using narrative categories as a template. 
For four social media events, we compared stories generated by Storia with stories that were crowdsourced without using narrative structure through a controlled study. We asked 30 participants to evaluate each story with respect to how well it conveys an event to someone who had missed it, and found that, for certain events, Storia stories were recommended three times as often by participants due to its emotional content.
%\todo{We then asked \todo{X} participants to evaluate both stories with respect to factual accuracy, trustworthiness, and emotional understanding of the event. Through an analysis of survey responses and interviews, we found that TODO}. 

To summarize our contributions, in this paper we:

\begin{itemize}
\itemsep0em
\item contribute a technique for recovering missing information from social media feeds by identifying and filling narrative gaps,
\item demonstrate the application of narrative theory in designing crowdsourcing workflows for generating stories, and
\item explore the limitations of narrative summarization by studying its output given different types of social media events.
\end{itemize}

Our results set the stage for constructing concise emotional experiences out of multiple viewpoints and deriving lessons for applying narrative theories to approaches for crowdsourcing creative work. 

%To see whether it is possible for the crowd to use these strategies to interpret and generate a coherent picture of an event based on social media, 
%Turning social media content into a story with emotional impact requires placing these major moments into a broader narrative arc.
%Effective authors of both fiction and non-fiction use \emph{narrative structure} to provide context and invoke empathy, which typically includes the introductions of characters and setting, an inciting problem that drives a plot, a central conflict, and a resolution \cite{chatman1980story}.
% Narrative theorists then seek to understand how to create narrative and what sets narrative apart from other forms of discourse. We can view social media content as artifacts of attempts to understand collective experiences; social media inherently structures itself around the relationship between a storyteller (the user or original poster) and an audience (other users, followers, friends).

\section{Related Work}
%Interest in the role of vizualizations in telling stories based on large amounts of user-generated content has increased \cite{Segel:2010:NVT:1907651.1908000}. However, 
Storia focuses on using an \emph{underutilized source of content} (that is, social media) to craft narratives, rather than surfacing trends and themes for analysis or (re)constructing a logical description of events. It turns to past work in narrative theory, social media summarization and curation, and crowdsourcing creativity to inform its design.

\subsection{Narrative Theory}
Narrative theory \cite{gibson1996towards} stems from the idea that people use narrative as a basic cognitive strategy for making sense of various aspects of the human experience (such as time and change).
%Narrative theorists then seek to understand how to create narrative and what sets narrative apart from other forms of discourse, and view the study of narrative as an important window into human cognition. 
In fact, the presence of narrative can significantly alter how an experience is perceived \cite{schneider2004death}. We may be able to frame social media content as the product of people attempting to understand experiences with others; when a user creates and posts content, they act as a narrator conveying some experience to an audience.
%The major design hypothesis underlying the crowdsourcing tasks that comprise Storia is that the miniature narratives created by individuals can together form a larger, coherent narrative of an event or topic.

Storia attempts to incorporate specific theories about how stories are understood in the design of its crowdsourcing tasks. In his theory of narrative categories, Cohn approaches storytelling in comics with respect to cognition, examining visual elements in terms of narrative syntax \cite{cohn2013visual}. Cohn argues that individual comic panels can be mapped to four basic narrative functions or roles that control narrative dramatization and pacing:
\begin{itemize}
\itemsep0em
\item the \emph{peak} depicts the culmination of an action set in motion during the narration; it can stand alone as a (blunt) summary of the narrative. 
\item the \emph{establisher} sets up the relationships of all characters involved in the story, 
\item the \emph{initial} starts the action that eventually culminates in the peak, 
\item and the \emph{release} depicts the aftermath or reaction to the peak, providing a sense of closure or creating room for anticipation for the next part of the story. 
\end{itemize}

Furthermore, these roles can act hierarchically: a group of panels can together fulfill a narrative role for the larger story. In other words, these roles form ``sentences'' that make up the narrative arcs of a comic. Storia uses this theory as the basis for a structured form used by crowd workers to write text (rather than visual) summaries based on social media data.

\subsection{Social Media Summarization and Curation}
Projects such as Narrative Science \cite{narrativescience} point to the value of transforming large amounts of quantitative data into natural language summaries to facilitate an accessible understanding of an event. A large body of work focuses specifically on text summarization of events in social media, particularly on Twitter \cite{Nichols:2012:SSE:2166966.2166999,chakrabarti2011event}.
%Sharifi:2010:SMA:1857999.1858099}.
This work focuses mostly on using text analysis, sentiment analysis, and machine learning \cite{Marcus:2011:TAV:1978942.1978975,5652922, ICWSM148117} to surface important moments out of social media chatter and on generating understandable summary text automatically. Other work aggregates Twitter content into visual summaries \cite{o2010tweetmotif,Shamma:2011:PPM:1958824.1958878}.
%Hannon:2011:PAS:1943403.1943459,Shamma:2009:TDU:1631144.1631148,}.
However, while visual summaries may \emph{suggest} a narrative, they do not offer a narrative on their own. Both types of summaries often result in a simple list of highly-tweeted moments (such as the goals in a soccer game) that is detached from the emotional ups and downs of the overall event. We complement this past work; rather than try to detect important events, we start with a data set that has already grouped social media content into important moments using existing techniques and attempt to form a \emph{narrative} that provides the context behind various pieces of information and conveys a sense of dramatic structure. 

%Past work with similar goals have used sentiment analysis to create a picture of the emotional landscape of an event as seen by social media; this body of work tends to analyze user content in terms of positive and negative sentiment \cite{Marcus:2011:TAV:1978942.1978975,zhao2011analyzing,5652922}. EmotionWatch \cite{ICWSM148117} notably classifies Twitter posts into more fine-grained emotional categories and demonstrates how to visualize changes in emotions in relation to an event over time. While EmotionWatch focuses on creating a visual representation to aid in emotional analysis rather than generating a story, its use of time suggests that narrative summaries of an event may be valuable.

In social media curation, people \emph{manually} organize social media content in order to engage in sensemaking---for example, by collecting posts from a specific conversation or event. A major aspect of research studying social media curation involves developing systems that can assist human curators in sorting through large amounts of content, often by providing automatic techniques for recommending new and relevant content \cite{ICWSM124578,zhong2013sharing}.
%Tools for creating curated content such as Storify [ref] often allow users to annotate their collections with interpretations or explanations, and the problem of sorting through large amounts of content to find the right content to curate is a difficult one and have been addressed through automatic approaches that focus on recommendation . 
We extend this approach; in this paper, machines assist by providing \emph{structure} rather than direct suggestions, and crowd workers then interpret the provided structure to generate new content.
% In this work, workers go one step further---rather than selecting whole pieces of content as-is, they incorporate content together, surfacing the information they think is most important through the summaries they write (not unlike . Furthermore, multiple 

\subsection{Crowdsourcing Stories and Reports}
Crowdsourcing is often used as a tool to generate ideas and break down creative tasks into smaller pieces \cite{Kittur:2013:FCW:2441776.2441923}. In Newspad \cite{Matias:2014:NDC:2611780.2581354} and Eventful \cite{export:211606}, a story author asks crowd workers to create or retrieve content for news stories---for example, by asking workers to go to a particular event and take a specific set of photos. By delegating work to multiple people and drawing from content that has been already created, an author can more quickly collect a diverse set of content than if they were to work alone.

The application of crowdsourcing to more artistic work is also an emerging area of research. For example, in Ensemble \cite{Kim:2014:EEC:2531602.2531638}, authors maintained an outline of a short fiction story to guide crowd workers in generating ideas and contributing content.
%Pipeline \cite{Luther:2013:RLO:2441776.2441891} demonstrated that leaders may be able to redistribute creative responsibilities to others without compromising the success of a collaborative art project. 
While machines have showed promise in their ability to logically reorder or generate content according to narrative templates and structures, Storia hypothesizes we can augment this past work by making use of people's unique capacity to understand and create emotion.

\section{Storia}
Storia is a system that generates summaries of social media events through crowdsourcing tasks designed based on narrative theory. 
% Specifically, it comprises three crowdsourcing modules that generate summary paragraphs based on raw social media data.

\begin{figure}[!t]
\centering
\includegraphics[width=1\columnwidth]{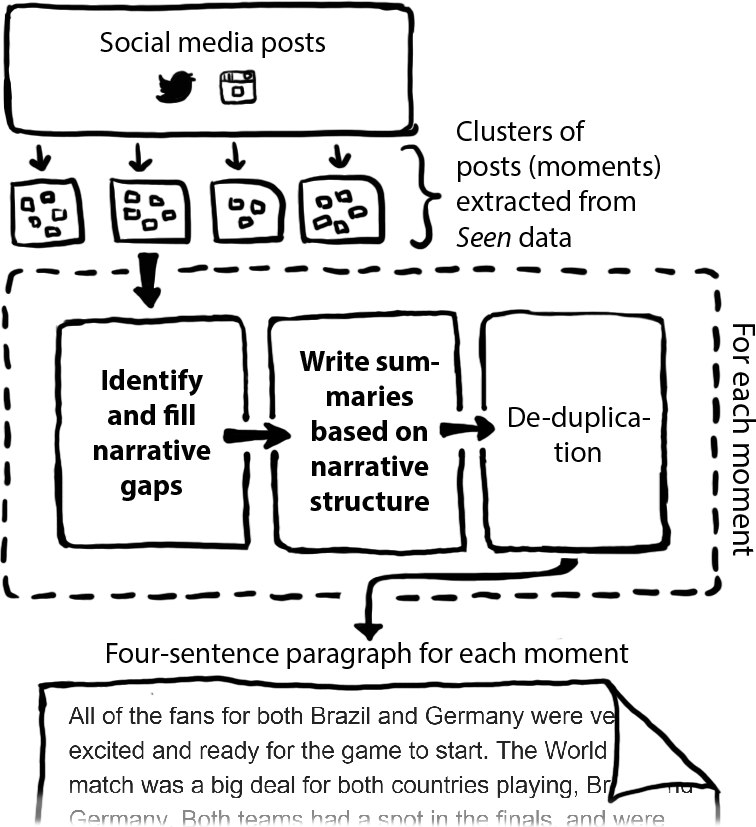}
\caption{Overview of the Storia system.}
\label{fig:simple-data-driven-comics}
\end{figure}

\subsection{Data}
Storia is based on content from \emph{Seen}\footnote{\url{http://seen.co/about}}, an online service that creates clusters of social media posts from Instagram, Twitter, and Vine based on time and keyword given a social media hashtag. Each cluster represents trending sub-issues or points in time related to the hashtag (e.g. tweets about the first goal of the \#GERvsARG soccer game). In this paper, we call each cluster of posts a \emph{moment}; several moments make up an \emph{event}.

%\subsection{Preliminary Interviews}

%Participants were between the ages of 20 and 39, and their jobs ranged from student to homemaker to firefighter. No participant dealt with social media as part of their job (for example, as a public relations manager) and all participants were regular users of social media networks such as Twitter or Instagram. During the in-lab study, participants were randomly assigned to one of the following views of a different set of a data (the 2014 FIFA World Cup final match between Germany and Argentina):
%\begin{itemize}
%\item \emph{Feed}: a simple chronologically ordered list of social media posts from the \emph{Seen} data.
%\item \emph{Timeline}: social media posts are displayed in clusters of time and by keyword. Clusters (or moments) are arranged in chronological order.
%\item \emph{News}: an article \cite{finalnews} from the \emph{New York Times} about the event. 
%\item \emph{Structured Summaries}: a set of summary paragraphs generated about the event by an early prototype of our workflow.
%\end{itemize}

%Participants were then asked to answer a questionnaire about factual and emotional aspects of the event. The questionnaire was designed to probe for how participants look for information about an event and asked questions such as ``Which Argentinean player made an offside shot?'' and ``How did viewers feel after the game ended?''. Finally, we conducted semi-structured interviews with participants, asking about typical 

\subsection{Design Challenge: Narrative Gaps}
To understand the role social media play in how people construct a story about an event, we conducted preliminary interviews with 10 participants (six male, four female) recruited from research volunteer mailing lists.
%who were each given \$100 in gifts to participate. 
We showed participants a variety of views of the 2014 FIFA World Cup final match between Germany and Argentina (a raw social media feed, a timeline of clustered social media posts, and a news article) and observed them as they used the views to learn about factual and emotional aspects of the event. Afterwards, we asked about the strategies they use for finding information about events, the role social media play in these strategies, perceptions of the social media view they were given, and motivations behind sharing information and social media posts with others. Each session lasted about 45 minutes.

Through these interviews, we found three common themes in the type of information participants looked for while attempting to construct a picture of the event:

First, there was a lack of understanding of the \emph{relationship between moments} during the event; even when social media posts were divided into clusters, participants had difficulty identifying the discrete parts that made up an event and how those parts related to one another:

\begin{quote}
\emph{I was trying to figure out the order of events like when that damn free kick happened. I have no idea.}

Participant 5, timeline condition
\end{quote}

Some participants had an easier time making predictions about where information might be located based on the fact that the story views shown to participants presented information in rough chronological order:

\begin{quote}
\emph{So I'm kind of like, from when everybody was like, ``Oh, Germany won.'' And just going right back from there and looking for ``free kick'' and a name in the post.}

Participant 3, timeline condition
\end{quote}

However, this strategy required \emph{domain knowledge} to be useful; the participant above knew that free kicks usually happen near the end of a soccer match. 

Second, participants were also confused about the \emph{relationship between actors}. When asked whether Brazilians generally rooted for Argentina or Germany during the final match, eight participants stated they didn't know. (Brazilians cheered for Germany, as Argentina is traditionally their rival with respect to soccer.) Thus, the potential conflict and outcomes at stake for each of the event's ``characters'' were not clear to participants.

Lastly, most participants saw no \emph{relationship between the event and their own lives}. Participants expressed little motivation to share or act on information found through social media with friends and family if they were not already interested or invested in the event in some way. We note that this is not necessarily an intrinsic weakness of social media but was also affected by participants' individual interests; nevertheless, participants indicated that emotional investment was necessary for them to take further action regarding the event.

It is unsurprising that participants had difficulty finding this information (even when the information they were looking for was available in the data). Social media posts are generally made in the moment \cite{Schirra:2014:TAM:2611205.2557070}, and so may make little sense to future readers; these posts might refer to people using pronouns or nicknames, or simply comment on a moment without describing what they are referring to (``Did you see that!?!?''). Furthermore, finding information is a learned skill, and the difficulties people encountered may not be solely inherent to social media; supporting information retrieval and search is a large research area on its own \cite{Marchionini:2006:ESF:1121949.1121979}. Given these difficulties, automating the creation of narratives may allow people to understand large amounts of data more easily.

\subsection{Extracting Narrative from Social Media}
With these challenges in mind, we developed a prototype story creation system called Storia (Figure \ref{fig:simple-data-driven-comics}) comprised of three crowdsourcing modules. Storia takes as input a set of raw social media content clustered by moment, and outputs a written story consisting of several four-sentence paragraphs (one paragraph per moment). 

Storia uses Amazon Mechanical Turk, an online crowdsourcing platform where workers can perform short microtasks for pay. In all tasks, crowd workers were from the U.S. who had a Mechanical Turk approval rating of over 90\%. 

\begin{figure}[!tb]
\centering
\includegraphics[width=1\columnwidth]{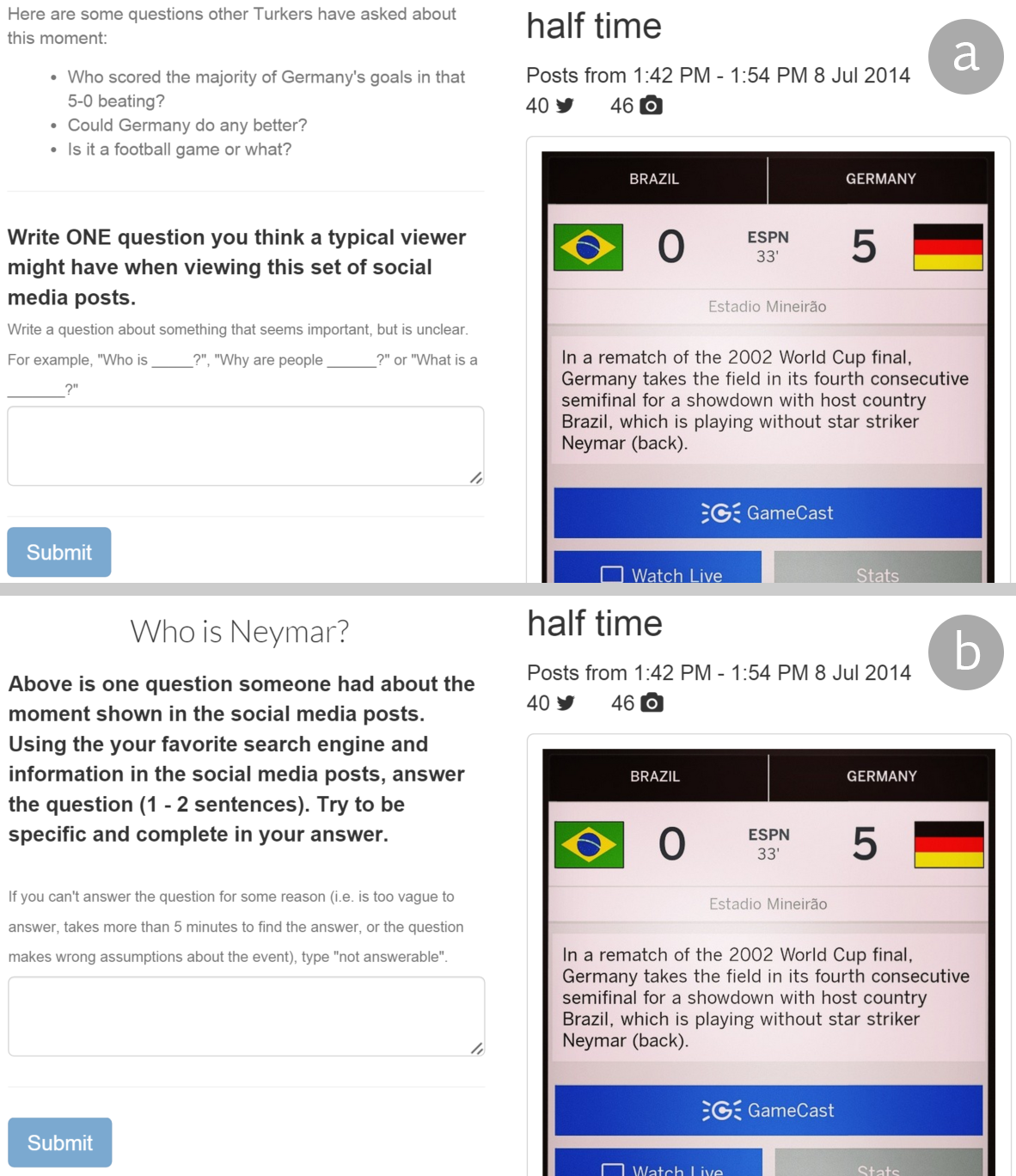}
\caption{Crowd workers can (a) view the social media posts for the moment they are assigned to and ask a question that will be (b) answered by another crowd worker.}
\label{fig:narrative_gaps_figure}
\end{figure}

\begin{figure}[!tb]
\centering
\includegraphics[width=1\columnwidth]{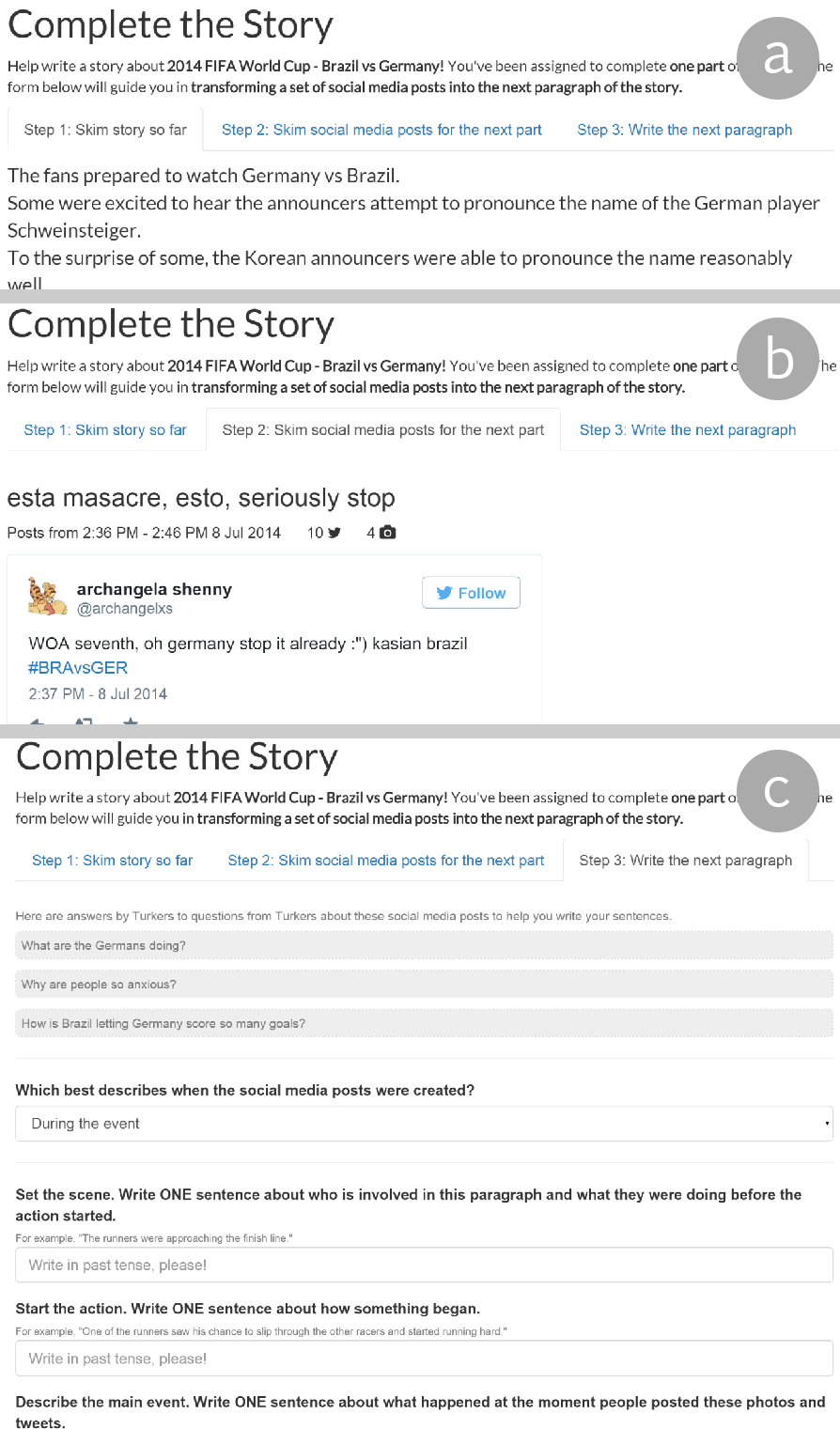}
\caption{Storia guides summary writing for crowd workers. Workers are provided with (a) the story written so far, (b) the social media posts for a moment of the event, and (c) a structured form with prompts based on narrative categories.}
\label{fig:figure_summary}
\end{figure}

\subsubsection{Identifying and Filling Narrative Gaps}

We first eliminate the narrative gaps we observed in our preliminary interviews. To alleviate confusion workers might have about unfamiliar names or terms encountered while processing social media content, we wanted to add context to the characters and actions that might appear in the final story. Two Mechanical Turk tasks (Figure \ref{fig:narrative_gaps_figure}) collected this information for each moment:

\emph{Ask questions.} For this task, we showed crowd workers a chronologically ordered stream of social media posts for a randomly chosen moment and asked workers what questions they thought a typical viewer might ask when viewing this content. Rather than suggest types of questions to ask based on narrative structure, we gave workers free rein to ask any question relevant to the goal of understanding the moment represented by the social media feed. We wanted to be open to the possibility that workers might ask unexpected types of questions, but workers did tend to ask questions related to narrative elements such as character and progression of plot, such as questions about names (``Who is Oscar?''), jargon specific to the event (``What is a free kick?''), and why some action was occurring (``Why was everyone booing Fred?''), reflecting the types of narrative gaps we observed in preliminary interviews.

\emph{Answer questions.} Here, we showed workers a question created by a worker from the previous task and asked them to briefly answer the question using a search engine or the information present in the social media stream. These workers saw the same stream of social media shown to the worker who had asked the question.

We collected two or three questions for each moment, and collected at least one answer for each question.

\subsubsection{Writing Summaries using Narrative Categories}

Second, we link content to narrative roles to generate a story structure for the event. In this module, we asked crowd workers to write one four-sentence paragraph for each moment of the event. Storia utilizes Cohn's four basic narrative categories (establisher, initial, peak, release) as a simple narrative template that crowd workers use as a base for the paragraph they write. Workers were provided with a view of paragraphs written by other workers for the overall story so far (Figure \ref{fig:figure_summary}a), the set of social media posts about the moment for which they were writing their paragraph (Figure \ref{fig:figure_summary}b), and the set of questions and answers generated by workers in the previous step. A structured form (Figure \ref{fig:figure_summary}c) prompted workers in mapping the information available in the social media feed to each of the Cohn's basic narrative categories. The wording of the prompts used in the form differed slightly (Table~\ref{tab:prompts_table}) depending on when the worker thought the moment occurred (i.e., near the beginning, middle, or end of the event). This process resulted in a story consisting of several paragraphs (one paragraph per moment), where each paragraph was composed of four sentences that map to each of Cohn's four basic narrative categories.

\renewcommand{\tabularxcolumn}[1]{>{\small}m{#1}}
\begin{table*}[!t]
  \centering
  \begin{tabularx}{\textwidth}{c X X X X}
    % & \multicolumn{4}{c}{\emph{Narrative Category}}
    %\tabularnewline
    \toprule
     &
    \multicolumn{1}{c}{\textbf{Establisher}} & 
    \multicolumn{1}{c}{\textbf{Initial}} & 
    \multicolumn{1}{c}{\textbf{Peak}} & 
    \multicolumn{1}{c}{\textbf{Release}}
    \tabularnewline
    \midrule
    \textbf{Beginning} & Make an introduction. Write ONE sentence about anyone or anything that hasn't appeared in the story yet, with a brief description about who/what they are. & Provide context. Write ONE sentence briefly summarizing the event as a whole. & Describe goals. Write ONE sentence about what the people or things you wrote above want to do during this event. & Describe the stakes. Write ONE sentence about what it would mean if they achieve their goal (or if they don't). \\
    \hline
    \textbf{Middle} & Set the scene. Write ONE sentence about who is involved in this paragraph and what they were doing before the action started. & Start the action. Write ONE sentence about how something began. & Describe the main event. Write ONE sentence about what happened at the moment people posted these photos and tweets. & Resolve the action. Write ONE sentence describing the aftermath or the reaction to the moment you see in these photos and tweets. \\
    \hline
    \textbf{End} & Summarize the result. Write ONE sentence about the important ending result of the event as a whole. & Summarize the reaction. Write ONE sentence about how people reacted to the result you wrote about in the sentence above. & Describe the consequences. Write ONE sentence about what the end result of the event means for the future. & Look to the future. Write ONE sentence about the next event or the next goal for the characters in this story. \\
    \bottomrule
  \end{tabularx}
  \caption{Prompts guided summary writing by workers. Storia prompts changed depending on whether the worker thought a moment occured at the beginning, middle, or end of the event.}
  \label{tab:prompts_table}
\end{table*}

We collected at least three paragraphs per moment then launched another task to ask other workers to vote for the paragraph they thought best represented the set of social media posts belonging to the moment. The highest voted paragraph became the representative paragraph for that moment. Votes were weighted slightly more if the voting worker indicated that he or she had watched the event.

\subsubsection{De-duplication}

Once we had one paragraph per moment, we ran a redundancy elimination task on Mechanical Turk. This is because the structured data we received from Seen.co did not produce mutually exclusive clusters---sometimes content appeared in multiple clusters, making it possible for the same content to inform multiple summaries by workers.

We showed workers a random paragraph from the story and asked them to select other paragraphs in the story they thought could replace their assigned paragraph without drastically affecting the story's overall flow or the information available to a reader. Workers were then asked to vote for the paragraph that best represented the entire group of paragraphs that they had selected.

We then grouped paragraphs by similarity. If at least two workers indicated that they thought Paragraph A was similar to Paragraph B, we considered Paragraph A and Paragraph B as true duplicates. If one of the two duplicate paragraphs was already in a group, we simply added the other paragraph to the same group; similarly, if both paragraphs were already in groups, their respective groups were combined. We then tallied the votes for all paragraphs and used the highest-voted paragraph from each group for the final story.

\section{Evaluation} % Results are here too
We hypothesized that recognizing and organizing social media data according to narrative roles could help workers overcome narrative gaps present in social media in order to produce evocative and automatic summaries of social media events.
%result in stories that help readers understand the emotional arc of an event. In this section, we investigate whether this strategy of structuring crowdwork using narrative theories produces narratives that change how people understand social media events.

\subsection{Method}
We tested our hypothesis through a controlled study comparing the output of two crowdsourcing workflows: the workflow used by Storia, described above (see Appendix A), and a control version of the workflow where we asked crowd workers to write four-sentence paragraphs for each moment of the event, but without prompts to guide the summary writing phase or the questions and answers generated about the event (see Appendix B). All other aspects of the control workflow remained the same as in the Storia workflow: paragraphs in the control condition went through a de-duplication process similar to that of the Storia condition in order to create a story with one paragraph for each moment.

We ran both the Storia and control workflows over social media posts about four different events (Table~\ref{tab:story_table}), and randomly sampled up to 12,000 posts from the entire corpus of content for each event. Stories ranged from sports events to television specials; in all cases, we chose widely viewed events based on topics that most crowd workers would be familiar with. Furthermore, we chose events that were well-structured (making it easy to compare crowd-created interpretations of the event with the actual sequence of events) and had some element of emotional arousal (and thus suitable for narration rather than just description). 
%1483 social media posts clustered into 28 moments were retrieved from Seen.co; each moment consisted of a mean of 53 posts ($SD = 32.35$).
%We chose this event as a source of data because it was a widely viewed event that generated lots of social media content and because it dealt with a broad topic most crowd workers would be somewhat familiar with (sports and soccer). Furthermore, sports events are both well-structured (making it easy to compare crowd-created interpretations of the event with the actual sequence of events) and have some element of emotional arousal (and thus suitable for narration rather than just description). 
We may be able to apply our findings to events that are interpreted and processed over an indeterminate amount of time (such as natural disasters and other breaking news), but leave this as future work.

To measure how well a non-expert reader might understand the generated text stories for each condition, we asked 30 Mechanical Turk workers to complete a short evaluation task. As we were interested in emotional reactions of a general population rather than an objective sense of the quality of stories written, we chose not to evaluate stories with experts. Participants were shown the Storia and control stories for a random event in random order, then asked to choose the story they would be more likely to recommend to someone who had wanted to see the event but missed it. We also asked workers to briefly explain their choice. Participants were paid \$0.30 for this two-minute task. Participants' free-form responses to the task were analyzed to look for themes in how participants justified their choice.

A second within-subjects evaluation task asked 30 additional Mechanical Turk workers to evaluate stories according to more fine-grained dimensions (such as informativeness) using 7-point Likert scales. The stories from both conditions for a randomly chosen event were shown in random order. Participants were paid \$0.40 for this task. For all evaluation tasks, participants were restricted to Mechanical Turk workers who had not participated in any of the story creation tasks.

%Lastly, we analyzed the content of both stories \todo{through an open coding process} to understand how crowd workers used (or mis-used) facts and descriptions of emotions and action.

\begin{table}[!t]
  \centering
  \begin{tabularx}{\columnwidth}{X c c c}
    \toprule
    \multicolumn{1}{c}{\textbf{Event}} &
    \multicolumn{1}{c}{\textbf{\# posts}} & 
    \multicolumn{1}{c}{\textbf{\# moments}} & 
    \multicolumn{1}{c}{\textbf{Event Date}} \\
    \midrule
    Sochi Winter Olympics Opening Ceremony & 4691 & 27 & 7 Feb. 2014 \\
    \hline
    2014 FIFA World Cup Semi-finals & 1483 & 45 & 8 Jul. 2014 \\
    \hline
    State of the Union (SOTU) 2015 & 11921 & 48 & 20 Jan. 2015 \\
    %\hline
    %Superbowl XLIX & 14340 & 153 & 1 Feb. 2015 & \todo{X} & \todo{X} \\
    %\hline
    %The Oscars -- 87th Academy Awards & 12822 & 339 & 22 Feb. 2015 & \todo{X} & \todo{X} \\
    \hline
    Glee Series Finale & 5574 & 29 & 20 Mar. 2015 \\
    \bottomrule
  \end{tabularx}
  \caption{The events used to generate stories through Storia and the control system. Content was randomly sampled from the entire corpus of posts for each event.}
  \label{tab:story_table}
\end{table}

\begin{table}[!t]
  \centering
  \begin{tabularx}{\columnwidth}{X c c}
    \toprule
    \multicolumn{1}{c}{\textbf{Task}} &
    \textbf{\# of HITs} & 
    \textbf{\$ per HIT} \\
    \midrule
    \normalsize Ask questions & 2 per moment & \$0.10 \\
    \hline
    \normalsize Answer questions & 2 per question & \$0.20 \\
    \hline
    \normalsize Write summaries & 3 per moment & \$0.50 \\
    \hline
    \normalsize Voting for summaries & 5 per moment & \$0.15 \\
    \hline
    \normalsize De-duplication & $\geq$ 2 per moment & \$0.30 \\
    \bottomrule
  \end{tabularx}
  \caption{Summary of the tasks and costs for both the Storia workflow and the control workflow.}
  \label{tab:final_table}
\end{table}

\subsection{Results}
Event type seemed to have a strong effect on the differences we observed between the Storia and control stories for each event. For this reason, we divide this section into two parts --- the first section describes the results for the FIFA and Winter Olympics events, and the second section describes the results for the SOTU and Glee events.

\subsubsection{``Placing me back in the game''}
For the FIFA and Winter Olympics events, crowd workers followed the establisher-initial-peak-release pattern in the paragraphs written for Storia stories, as seen in this example paragraph from the FIFA semi-finals Storia story:

\begin{quote}
\emph{The fans are sitting in front of their TVs and smartphones getting excited about the match as it starts.}

\emph{Germany scores their first goal against Brazil, and the fans are going wild rooting for Germany.}

\emph{Germany then goes on to scored their second, then their third and finally their fourth goal against Brazil, who has zero goals.}

\emph{Fans cannot believe what they're seeing and they're wondering if this is a match or a bloodbath because Germany has completely demolished Brazil.}
\end{quote}

In contrast, paragraphs written for control stories conveyed less of a dramatic arc, and instead tended to dwell on the same idea for most of its sentences. For example, each sentence in the following paragraph from the Winter Olympics control story mentions that viewers are ready for the event: 

\begin{quote}
\emph{People watching the ceremony announced they were ready for it to begin.}

\emph{The people watching were ready to support their individual nations.}

\emph{People tweeted out picture of themselves wearing gear showing their commitment to their country.}

\emph{Some people even tweeted out pictures of babies getting ready for their first opening ceremony.}
\end{quote}

%For example, each sentence mentions that the game has ended with Germany as the victor:

%\begin{quote}
%\emph{After thirty minutes of the Germany-Brasil game, the game was over.}

%\emph{With a score of 0-5 Germany fans were already celebrating.}

%\emph{While some felt bad for Brasil fans, most were happy with the results.}

%\emph{German fans were ecstatic that the chance to play in the World Cup final was given so easily to them.}
%\end{quote}

%Notably, we found that after de-duplication, workers generated 15 paragraphs for the story in the Storia condition, and generated 6 paragraphs for the story in the control condition, suggesting that workers seemed to find the control paragraphs written for each moment of the event more repetitive than the paragraphs written for the Storia story, resulting in less unique paragraphs for the control condition. 
For these two events, participants significantly preferred Storia stories over control stories (FIFA: $\chi^2(1)=6.5333, p<0.05$; Winter Olympics: $\chi^2(1)=10.8, p<0.01$); participants preferred the Storia story over the control story 73\% of the time for the FIFA event and 80\% of the time for the Winter Olympics event.
%\todo{Furthermore, there was a significant effect of study condition on how participants rated their emotional arousal after reading each story, with participants giving the Storia story an emotional rating of X ($SD=X$) and the control story an emotional rating of Y ($SD=Y$).}

Participants who picked Storia stories appreciated the large amount of detail included and felt that they were a more complete view of the event. Notably, most of the participants also justified their choice with some variant of ``I felt like I was getting a vivid recap'' or ``the story captured the emotion'':

\begin{quote}
\emph{While it would be simple to just say that [the Storia story] is longer etc...  It actually really expresses more emotion, more detail, and the ability to get a real feel for how the game went, the sentiments involved, everything to make it a better read!}

Participant, FIFA event
\end{quote}

Participants who voted for the control story stated they chose it because it was more concise, conveying major points about the event without including extraneous information:

\begin{quote}
\emph{[The Storia story] seems like a lot of non-quality information designed to entertain... while [the control story] is more informative.}
%\emph{[The control story] was much more concise and still got the same message across that [the Storia story] was trying to convey.}

Participant, Winter Olympics event
\end{quote}

To these participants, the control story seemed more professional. However, the participants that chose Storia stories stated they \emph{did not} pick control stories for very similar reasons: the control story felt like a bland generalization or a brief report. The second set of participants corroborated this sentiment; Friedman tests indicated that participants thought Storia stories had more interesting introductions (FIFA: $\chi^2(1)=8.067, p<0.01$; Winter Olympics: $\chi^2(1)=5.762, p<0.05$), giving Storia stories mean scores of 4.679 ($SD=1.307$) for the FIFA event and 5.111 ($SD=1.22$) for the Winter Olympics event, and control stories mean scores of 3.571 ($SD=1.501$) for the FIFA event and 4.37 ($SD=1.363$) for the Winter Olympics event.

Participants also thought Storia stories for these events were more informative (FIFA: $\chi^2(1)=16.2, p<0.01$; Winter Olympics: $\chi^2(1)=7.1176, p<0.01$), giving Storia stories mean scores of 5.679 ($SD=0.612$) for the FIFA event and 5.519 ($SD=1.087$) for the Winter Olympics event, and control stories mean scores of 4.393 ($SD=1.343$) for the FIFA event and 4.778 ($SD=1.086$) for the Winter Olympics event.

Lastly, participants felt that the Storia stories for these events were more likely to make readers feel as if they were there (FIFA: $\chi^2(1)=15.385, p<0.01$; Winter Olympics: $\chi^2(1)=4.262, p<0.05$), giving Storia stories mean scores of 5.143 ($SD=1.079$) for the FIFA event and 5 ($SD=1.144$) for the Winter Olympics event, and control stories mean scores of 3.25 ($SD=1.404$) for the FIFA event and 4.111 ($SD=1.528$) for the Winter Olympics event.

% Whoops climax isn't true anymore with new data
%Participants also thought the Storia had a more exciting climax ($\chi^2(1, N = 30) = 3.8571, p < 0.05$), giving the Storia story a mean score of 4.9 ($SD = 1.16$) and the control story a mean score of 3.97 ($SD = 1.35$).

\subsubsection{``I disliked both stories, but...''}
The stories for the Glee and SOTU events were characteristically different from the stories for the FIFA and Winter Olympics events. For example, in both SOTU stories, workers added their own opinions regarding certain politicians using words not present in the provided set of social media content:

\begin{quote}
\emph{President Obama walked into the SOTU with the normal pomp and circumstance.}

\emph{President Obama came out firm, expressing his overwhelming victories to the obstinate and denialist Congress.}

\emph{John Boehner had the same, half asleep, half drunk look on his face.}

\emph{With that, the gauntlet was thrown.}

\end{quote}

The control condition went so far to parody political relationships to the point of absurdity, which extended through several paragraphs:

\begin{quote}
\emph{Vladimir Putin was not happy, the endless enlargement of NATO could not stand!}

\emph{John McCain on the other hand couldn't help but peer over at his sore buddy Putin and laugh to himself.}

\emph{But the murmering in the crowd quieted as Barack Obama approached the podium...}

\emph{``Suck it Putin!'' he exclaimed as he ripped off his shirt and exposed the ``Superman S'' on his undershirt! ``I'm Barack Obama, you got that'' and he flew away.}

\end{quote}

For these stories, there was no significant effect of study condition on participants' preferences (Glee: $\chi^2(1)=0.133, n.s.$; SOTU: $\chi^2(1)=1.2, n.s.$). For the Glee story, most participants stated they based their choice on writing quality rather than on emotional aspects. The second set of participants reflected this, as there was no significant difference in which story they thought would be more likely to make readers feel like they were at the event. In hindsight, this makes sense, as these events were meant to be televised rather than attended.

Opinions on which story participants preferred for the SOTU event was divided on the control story's use of parody---some workers thought it was amusing, but others disapproved:

\begin{quote}
\emph{[The control story] was juvenile and unintelligent in too many places... [The Storia story] provides a better summary that is more sophisticated and intelligent (even though it's not great either).}

Participant, SOTU event
\end{quote}

This is reflected in the results from the second evaluation task; neither Glee story was seen as more informative than the other ($\chi^2(1)=7.118, n.s.$), but the Storia story for the SOTU event was seen as more informative ($\chi^2(1)=10.889, p<0.01$), receiving a mean score of 4.96 ($SD=1.695$) while the control story received a mean score of 3.28 ($SD=1.969$). Neither story, in both events, was seen as having a more interesting introduction (Glee: $\chi^2(1)=3.556, n.s.$; SOTU: $\chi^2(1)=2, n.s.$).

Overall, the approach of framing social media summarization around narrative seemed to be more effective for the FIFA and Winter Olympics events. For these events, participants did perceive the Storia story as having higher emotional value, choosing to recommend Storia stories to someone who wanted to learn more about the event.

\section{Discussion}
Through an evaluation of narratives generated by crowdsourcing tasks based on narrative theory, we found that certain types of stories written with narrative guidance were more emotionally engaging and better suited for conveying an event to someone who had missed it. Other events did not benefit as clearly from narrative guidance. Here, we discuss the strengths and limitations of this approach as seen in our results, as well as potential broader impact on future approaches for crowdsourcing creativity.

\subsection{Interpreting Social Media}
Notably, study participants did not observe emotional differences between the Storia and control stories for the Glee and SOTU events.
We speculate this is because the narrative gaps in these events were too large---in the Glee event's case, workers had little information about the overall arc of the finale episode, making it difficult to construct an emotional summary even with information retrieved by other crowd workers through questions and answers. Similarly, the SOTU event had no inherent narrative arc at all, with no clear winner, loser, or conflict. Because of this, the only meaningful thing workers could write about were their own opinions (or fictitious storylines). This is in contrast to sports events, which can be understandable and interesting even when one is not acquainted with the players or teams due to its familiar narrative and emotional pattern---a struggle to win, the joy of victory, and the pain of loss.

A strategy often employed by journalists for writing stories about events that do not have their own narrative arcs is to take a specific viewpoint.
%However, we may be able to constructively channel conflicting opinions to create new kinds of stories based on social media. 
Storia may be able to accommodate this strategy, as it does not provide deterministic output; depending on the workers who participate in the process,
%The fact that different stories can be generated for different perspectives may be Storia's greatest strength. 
one could view a State of the Union address from the perspective of a Democrat \emph{or} from the perspective a Republican. One could also view events from the perspective of a person watching the event at home, or from the perspective of the event organizers, or from the perspective of players or performers, each with their own goals and hopes for the event.
%However, the sheer amount of social media data available is difficult for an individual to parse. The modularity of Storia stories may be a good fit for supporting multiple points of view. One could imagine being able to ``choose your own adventure'' and examine an event through several different storylines. 

Extending this further, this might allow us to apply Storia's approach to other types of social media, such as question-answering sites (like Quora) or discussion sites (like reddit). Rather than generating a story about an event (which implies a narrative structure due to having a beginning and end), the crowd could, for example, identify multiple stances for opinion pieces based on an online discussion.
%In future work, the crowd might identify possible points of view and generate Storia stories based on each possibility. 
These perspectives could then be available to anyone wanting to learn about different interpretations of the event. 
Stories are not reproductions of reality but representations of it (whether consciously designed or not). We have other means of learning what happened in reality (e.g. through recordings); this paper tries to look at how we can expose and examine an audience's interpretation of what happened.

%At the same time, people may compose stories about data that do not necessarily exist in attempts to fill narrative gaps and make sense of existing content. 
%Additionally, crowd workers may inject their own opinions and biases into the story through the creative process. For example, in a test story we ran on social media posts about the 2015 State of the Union speech, we found that many Turkers injected their opinions about American political parties in their work (for example, by writing satirical summaries using politicians as characters in fictional situations). 
%In this paper, we selected an event that made it easy to verify the stories generated by crowd workers (i.e. by comparing output with news reports or recordings), but this may not always be possible. However, 

%Machines may be able to cluster/structure responses and information surrounding a discussion topic or a question, that the crowd could then develop into a coherent piece. Storia, however, currently does not deal with helping the crowd decide on a higher-level point for their output---unlike a sports game, where there is always a winner and a loser, opinion pieces and knowledge articles require their author to take some sort of stance in order to produce a coherent argument or explanation. This could be interesting future work to extend Storia's approach to accommodate this necessary part of collaboration.

\subsection{Converging Creative Goals}
Crowdsourcing creative work typically requires splitting the project into smaller tasks \emph{a priori} and stitching together the results of each sub-task to create the final product \cite{Kittur:2013:FCW:2441776.2441923}. As a result, the success of a creative task depends heavily on the design of its sub-tasks. While Storia follows this approach in the sense that an event is described through a set of individual paragraphs, Storia also allows the crowd to write sub-tasks for itself: the crowd is able to ask questions that other workers can answer, for example.

The idea of asking crowd workers to generate crowdsourcing tasks for themselves is not new \cite{Kulkarni:2012:CCW:2145204.2145354}; however, Storia points toward a strategy of allowing the crowd to \emph{identify gaps in understanding}, which, in turn, becomes work for other crowd workers. That is, it dynamically structures creative work around testing and iteration rather on predetermined sub-tasks that eventually merge together. Enabling this flexibility may be even more important for creative work (such as story writing), which may have no objective solution and require workers to converge on common creative goals.

%\subsection{Using Narrative Structure in Creative Work}
%\todo{We also wanted to see if the narratively structured summaries we generated could help create stories in media beyond just text. There is some precedence for this; comics journalism \cite{comicsjournalism} and newsgames \cite{Bogost:2010:NJP:1941873} are both examples of media that convey non-fiction stories---including topics such as the Holocaust \cite{spiegelman1986maus} or immigration issues \cite{papersplease}---using techniques typically used for fiction. These media also explore how to use digital technology and the web to make news accessible to a wider demographic \cite{Chen:2013:CCO:2468356.2468830,treanor2009newsgames}.}

%\todo{Because we generate summaries with narrative structure rather than just text, we may be able to use our summaries as a base for other media. Furthermore, social media are multimedia -- eventually, we hope to use this strategy to generate richer and emotionally relevant summaries, and leave this as future work.}

\subsection{Limitations}
As explained previously, we chose to apply Storia's approach to events that had a clear beginning and an end, in order to ensure we would be able to compare crowd-created interpretations of the event with what actually happened.
For this reason, we were able to use an approach where social media is automatically clustered into parts and the crowd linearly transforms these clusters into a story.
This approach might be less suitable for events concerning natural disasters, breaking news, and other ongoing, developing stories---Storia currently relies on a story structure that is static, and interpreting the importance or emotional valence of certain moments may be difficult as new information and public opinion develops. 
%A natural next question is to ask whether this approach of identifying narrative gaps and summarizing based on a machine-generated structure would work well for events concerning natural disasters, breaking news, and other ongoing, developing stories. Our exact approach may be difficult to use in these cases, as the structure of the story and gaps in the knowledge of the audience may change as the story develop and more information comes to light. On the other hand, Storia's approach may act as a useful framework for understanding and documenting changes in the knowledge of the social media network for developing events.

Despite the fact that participants appreciated Storia's level of detail, both Storia and control stories were prone to factual errors. 
%For example, some paragraphs mistakenly claimed a fact (such as the final score for a sports match) that was incorrect; as another example, correct facts would sometimes appear in the wrong order---one story describes a German player scoring the first goal of the game, even though the story had already recounted the first four goals of the match. 
However, in this paper, our goal was to enable scalable creation of evocative and experientially-oriented summaries of social media events rather than aim for factual accuracy; we suspect additional strategies (such as asking a human editor to proofread stories generated by our system) can effectively address these issues.

Storia also limited itself to producing text output, despite the large number of visual content that appeared in social media feeds shown to workers. Images can be effective tools for expressing difficult-to-describe or intangible things such as emotion or atmosphere; in the New York Times' article about Brazil's loss to Germany in the 2014 FIFA World Cup semi-finals \cite{semifinalnews}, photos and videos of stunned fans and grieving players accompany the text. Exploring how crowd workers can make use of multiple types of media while generating their stories is left as future work.

%We suspect these errors are not caused by presence of narrative structure, but are instead byproducts of the fact that the Storia workflow generates more unique and detailed paragraphs; the more paragraphs present in the story, the more likely crowd workers are to contradict one another. A future version of the Storia workflow (or a human editor) could fact-check details and detect inconsistencies to prevent these types of errors. 

%We note that Cohn's narrative categories include a number of categories other than the ones we explore here with Storia and also include the idea that these categories can function hierarchically or can be modified depending on the relationships among them. Storia is an initial exploration into how narrative theory can inform the design of social computing tasks. Integrating these more complex ideas is left as future work.

\section{Conclusion}
In this work, we explored the strategies for creating stories about events based on social media data that convey a public interpretation of an event. Our prototype, Storia, drew together narrative theories and crowdsourcing to create a system that generates collaborative creative work by \emph{finding and filling narrative gaps} and \emph{linking content to narrative roles}. Through a controlled study, we compared stories generated by Storia to stories crowdsourced with no narrative guidance, and found that, for certain events, participants found Storia stories more emotionally engaging and more appropriate for communicating what it felt like to view an event.

Are emotional stories better than objective ones? Journalists often struggle to maintain a balance between objectivity and emotion when writing their stories. In journalism studies, the ``emotionalization'' of stories is often associated with sensationalism and the decline in quality of journalistic stories \cite{pantti2010value}; on the other hand, a case study of Pulitzer Prize-winning articles revealed that winning stories rely heavily on emotional storytelling, using strategies such as anecdotal introductions and expressions of affect in order to draw attention to complex and important social and political topics \cite{wahl2013strategic}.

We did not attempt to compare stories generated by Storia with professionally written news articles and stories. Storia does not attempt to automatically generate Pulitzer-prize winning articles, nor does it try to solve the problem of distinguishing emotionality from tabloidism; it is obvious that our generated stories are nowhere near professional quality. However, motivated by evidence that emotional perspectives of events are valued, it takes a first step towards considering emotional and narrative arcs in the automatic generation of event summaries. People join social media networks to ask questions, read comments, and make connections---all to seek out what others think and how they feel. Social media is rich with declarations of emotions; by distilling the chatter of the social web, we may be able to bring out its voice.

%Though we only explored social media surrounding soccer matches, we anticipate that the approach behind Storia could be applicable to other kinds of events. Storia is based on \emph{Seen} data, which clusters social media using an algorithm based on text and proximity in time. Events that slowly unfold over time (such as unexpected disasters or other breaking news events) are reflected on and processed by people and do not have a clear end, and so may not be the best match for Storia. Instead, Storia could apply its process to any event that can be clustered into time-oriented moments. 

\section{Acknowledgments}
We would like to thank our colleagues at Microsoft Research, study participants, and Mechanical Turk workers for their valuable feedback and participation, as well as Neil Cohn for his insights on narrative theories. We also thank Mor Naaman and Seen.co for their help with their dataset. This material
is based upon work supported by the NSF GRFP under Grant No. DGE-114747.

\section{Appendix}
\subsection{A. Example Storia Story}
\begin{footnotesize}
The fans prepared to watch Germany vs Brazil.
Some were excited to hear the announcers attempt to pronounce the name of the German player Schweinsteiger.
To the surprise of some, the Korean announcers were able to pronounce the name reasonably well.
The fans were impressed at the ability of the announcers to pronounce Schweinsteiger.

All of the fans for both Brazil and Germany were very excited and ready for the game to start.
The World Cup match was a big deal for both countries playing, Brazil and Germany.
Both teams had a spot in the finals, and were very anxious to win the game to proceed.
Losing the game would mean they would have no chance at the World Cup.

Brazil and Germany had both worked very hard to get to this point, and only one would be able to continue.
The fans were going wild as the opening ceremonies began.
Both teams played very well.
Only one team actually won.

David Luiz held the jersey of his injured teammate Neymar.
The team began to sing the national anthem.
Their enthusiasm shone brightly as they sang the anthem like a war cry.
Fans speculated whether David Luiz could lead the team to victory as they geared up for the match.

The fans are sitting in front of their TVs and smartphones getting excited about the match as it starts.
Germany scores their first goal against Brazil, and the fans are going wild rooting for Germany.
Germany then goes on to scored their second, then their third and finally their fourth goal against Brazil, who has zero goals. 
Fans cannot believe what they're seeing and they're wondering if this is a match or a bloodbath because Germany has completely demolished Brazil.

Germany was on fire and scored goal after goal.
The Brazilian fans started to become angered while the Germans were more and more elated.
Anger turned to despair for Brazilan fans as Germany extended their lead by an incredible margin.
As loss seemed inevitible, the fans lamented the absence of Brazilian forward Neymar, who had been injured in a previous match.

The game was already heavily in Germany's favor.
German player Lahm made a good tackle on Marcelo, winning the ball fairly.
The fans were very impressed with Lahm's performance.
It was just one more show of dominance by Germany.

Germany's leading scorer Thomas Muller drove towards the goal.
Muller scored the first goal of the game, and the fans went wild.
The fans remarked that this was Mueller's tenth World Cup goal, an impressive feat. 
Unfortunately for Brazil, things were only going to get worse from here.

Brazil and Germany were locked in a tense World Cup game
Fans are in disbelief about what has happened so far.
Fans are stunned that Germany is dominating Brazil.
Fans were in utter disbelief at the way the game was playing out.

The German players looked to still have a full tank of gas half way into the game.
The Brazilians tried without luck to stop the German attack, but German technique was to hog the ball.
Again, German took a shot at the goal with Schurrle moving into the goal box!
One tweeter posted ''They look like they've got concrete boots on

Brazil and Germany played a fierce game of soccer.
Germany's goalkeeper blocked every one of Brazil's attempts to score a goal.
The game ended in emberrassment for Brazil as they ended the game with 0 points, due to Germany's defensive goalkeeper.
Germany has now advanced to the finals, leaving Brazillian fans fuming across the world.

With Brazil's weakness in depth, they had to play Fred as a striker.
He showed promise in qualifying leading up to the event but has really struggled in Brazil.
As he looked invisible on the pitch and Germany ran up the score, people wondered why he was even on the field.
After the 7-0 scoreline flashed, fans wondered how a nation such as Brazil can have such little depth.

Germany managed to defeat Brazil 8-0
The Germany fans were ecstatic, while the Brazilians were shocked 
Brazil, previously thought to be one of the best, will have this game rubbed in their faces until the end of time
Germany, boosted by their win, will be helped in the next stage

Germany destroyed Brazil in the World Cup semifinal.
Fans were astounded by the massacre.
Brazil will suffer a setback on the international soccer stage.
Germany looks to win in the final

With Germany up 7 to 0, it appeared Brazil would go scoreless in this World Cup semifinal.
With only minutes left, Brazilian player Oscar managed to score one goal.
Now the score was 7 to 1, so at least Brazil had put some points on the board.
Many fans thanked Oscar for helping give a small shred of respect to the Brazilian rout.
\end{footnotesize}

\subsection{B. Example Control Story}
\begin{footnotesize}
Neuer, Lahm, Boateng, Kroos, Schweinsteiger and M\"{u}ller comprised Germany's starting lineup in the semi-finals.
Dante made it to Brazil's semi-finals, playing his first game for the World Cup.
It had been speculated that Willian would replace Neymar, but it turned out to be Bernard instead.
Dante replaced Thiago Silva, making Brazil's semi-final lineup Fred, Oscar \& Bernard.

While casual fans though Brazil would win die hard soccer fans though Germany had the better team.
It wasn't surprising how Germany played, but it was how Brazil played.
For some reason it just seemed that Brazil didn't show up to start the game and things steamrolled.
Germany was the best team at this World Cup and deserved the title.

Thomas M\"{u}ller scored the first point in the semi-finals!
Left unguarded, he snuck the ball in through the corner.
The fans went wild, as Germany put heavy pressure on Brazil with the early point.
With an impressive record of scores and assists in World Cup games, M\"{u}ller has truly shown himself as a strong asset to Germany's team.

After thirty minutes of the Germany-Brasil game, the game was over.
With a score of 0-5 Germany fans were already celebrating.
While some felt bad for Brasil fans, most were happy with the results.
German fans were ecstatic that the chance to play in the World Cup final was given so easily to them.

Fans around the world had much to say about Brazil's disappointing performance in the 2014 FIFA World Cup.
Some took to social media sites like Twitter to bash the Brazilian team.
German fans were excited about their team and took every opportunity to make fun of Brazil.
Brazilians were upset about their team.

As Germany crushed, you had to feel sorry for Brazil.
Goal keeper Manuel Neuer was in great from, shutting down Brazil.
Neymar was sulking, mighty Brazil was in trouble.
Finally, deep in stoppage time, Brazil got some relief as they had scored.
\end{footnotesize}

%\footnotesize %8pt

% REFERENCES FORMAT
% References must be the same font size as other body text.
\bibliographystyle{SIGCHI-Reference-Format}
\bibliography{proceedings}


\begin{thebibliography}{00}

%%% the Web address.

\ifx \showCODEN    \undefined \def \showCODEN     #1{\unskip}     \fi
\ifx \showDOI      \undefined \def \showDOI       #1{{\tt DOI:}\penalty0{#1}\ }
  \fi
\ifx \showISBNx    \undefined \def \showISBNx     #1{\unskip}     \fi
\ifx \showISBNxiii \undefined \def \showISBNxiii  #1{\unskip}     \fi
\ifx \showISSN     \undefined \def \showISSN      #1{\unskip}     \fi
\ifx \showLCCN     \undefined \def \showLCCN      #1{\unskip}     \fi
\ifx \shownote     \undefined \def \shownote      #1{#1}          \fi
\ifx \showarticletitle \undefined \def \showarticletitle #1{#1}   \fi
\ifx \showURL      \undefined \def \showURL       #1{#1}          \fi

\bibitem{narrativescience}
Narrative Science.
\newblock
\newblock
\newblock
\shownote{\url{http://www.narrativescience.com/}.}


\bibitem{storify}
{Storify}.
\newblock
\newblock
\newblock
\shownote{\url{http://storify.com/}.}


\bibitem{export:211606}
{Elena Agapie} {and} {Andr{\'e}s Monroy-Hernandez}. 2014.
\newblock \showarticletitle{Eventful: Crowdsourcing Local News Reporting}.
  Collective Intelligence Conference.
\newblock
\showURL{%
\url{http://research.microsoft.com/apps/pubs/default.aspx?id=211606}}


\bibitem{semifinalnews}
{Sam Borden}. 2014.
\newblock \showarticletitle{{World Cup 2014: Host Brazil Stunned by Germany in
  Semifinal}}.
\newblock {\em The New York Times\/} (July 2014).
\newblock


\bibitem{chakrabarti2011event}
{Deepayan Chakrabarti} {and} {Kunal Punera}. 2011.
\newblock \showarticletitle{Event Summarization Using Tweets}. In {\em Proc.
  ICWSM} {\em (ICWSM)}.
\newblock


\bibitem{cohn2013visual}
{Neil Cohn}. 2013.
\newblock \showarticletitle{Visual narrative structure}.
\newblock {\em Cognitive science\/} {37}, 3 (2013), 413--452.
\newblock


\bibitem{5652922}
{N. Diakopoulos}, {M. Naaman}, {and} {F. Kivran-Swaine}. 2010.
\newblock \showarticletitle{Diamonds in the rough: Social media visual
  analytics for journalistic inquiry}. In {\em IEEE Symposium on VAST}.
  115--122.
\newblock


\bibitem{ICWSM124578}
{Kevin Duh}, {Tsutomu Hirao}, {Akisato Kimura}, {Katsuhiko Ishiguro}, {Tomoharu
  Iwata}, {and} {Ching-Man~Au Yeung}. 2012.
\newblock \showarticletitle{Creating Stories: Social Curation of Twitter
  Messages}. In {\em Proc. ICWSM} {\em (ICWSM)}.
\newblock


\bibitem{gibson1996towards}
{Andrew Gibson}. 1996.
\newblock {\em Towards a postmodern theory of narrative}.
\newblock Edinburgh University Press, Edinburgh.
\newblock


\bibitem{ICWSM148117}
{Renato Kempter}, {Valentina Sintsova}, {Claudiu Musat}, {and} {Pearl Pu}.
  2014.
\newblock \showarticletitle{EmotionWatch: Visualizing Fine-Grained Emotions in
  Event-Related Tweets}. In {\em Proc. ICWSM}.
\newblock


\bibitem{Kim:2014:EEC:2531602.2531638}
{Joy Kim}, {Justin Cheng}, {and} {Michael~S. Bernstein}. 2014.
\newblock \showarticletitle{Ensemble: Exploring Complementary Strengths of
  Leaders and Crowds in Creative Collaboration}. In {\em Proc. CSCW}. 745--755.
\newblock
\showISBNx{978-1-4503-2540-0}


\bibitem{Kittur:2013:FCW:2441776.2441923}
{Aniket Kittur}, {Jeffrey~V. Nickerson}, {Michael Bernstein}, {Elizabeth
  Gerber}, {Aaron Shaw}, {John Zimmerman}, {Matt Lease}, {and} {John Horton}.
  2013.
\newblock \showarticletitle{The Future of Crowd Work}. In {\em Proc. CSCW}.
  ACM, New York, NY, USA, 1301--1318.
\newblock
\showISBNx{978-1-4503-1331-5}


\bibitem{Kulkarni:2012:CCW:2145204.2145354}
{Anand Kulkarni}, {Matthew Can}, {and} {Bj\"{o}rn Hartmann}. 2012.
\newblock \showarticletitle{Collaboratively Crowdsourcing Workflows with
  Turkomatic}. In {\em Proc. CSCW}. ACM, New York, NY, USA, 1003--1012.
\newblock
\showISBNx{978-1-4503-1086-4}


\bibitem{Marchionini:2006:ESF:1121949.1121979}
{Gary Marchionini}. 2006.
\newblock \showarticletitle{Exploratory Search: From Finding to Understanding}.
\newblock {\em Commun. ACM\/} {49}, 4 (April 2006), 41--46.
\newblock
\showISSN{0001-0782}


\bibitem{Marcus:2011:TAV:1978942.1978975}
{Adam Marcus}, {Michael~S. Bernstein}, {Osama Badar}, {David~R. Karger},
  {Samuel Madden}, {and} {Robert~C. Miller}. 2011.
\newblock \showarticletitle{Twitinfo: Aggregating and Visualizing Microblogs
  for Event Exploration}. In {\em Proc. CHI}. ACM, New York, NY, USA, 227--236.
\newblock
\showISBNx{978-1-4503-0228-9}


\bibitem{Matias:2014:NDC:2611780.2581354}
{J.~Nathan Matias} {and} {Andr{\'e}s Monroy-Hernandez}. 2014.
\newblock \showarticletitle{NewsPad: Designing for Collaborative Storytelling
  in Neighborhoods}. In {\em Proc. CHI}. 1987--1992.
\newblock
\showISBNx{978-1-4503-2474-8}


\bibitem{Nichols:2012:SSE:2166966.2166999}
{Jeffrey Nichols}, {Jalal Mahmud}, {and} {Clemens Drews}. 2012.
\newblock \showarticletitle{Summarizing Sporting Events Using Twitter}. In {\em
  Proc. IUI}. ACM, New York, NY, USA, 189--198.
\newblock
\showISBNx{978-1-4503-1048-2}


\bibitem{o2010tweetmotif}
{Brendan O'Connor}, {Michel Krieger}, {and} {David Ahn}. 2010.
\newblock \showarticletitle{TweetMotif: Exploratory Search and Topic
  Summarization for Twitter}. In {\em Proc. ICWSM}.
\newblock


\bibitem{pantti2010value}
{Mervi Pantti}. 2010.
\newblock \showarticletitle{The value of emotion: An examination of television
  journalists’ notions on emotionality}.
\newblock {\em European Journal of Communication\/} {25}, 2 (2010), 168--181.
\newblock


\bibitem{Schirra:2014:TAM:2611205.2557070}
{Steven Schirra}, {Huan Sun}, {and} {Frank Bentley}. 2014.
\newblock \showarticletitle{Together Alone: Motivations for Live-tweeting a
  Television Series}. In {\em Proc. CHI}. 2441--2450.
\newblock
\showISBNx{978-1-4503-2473-1}


\bibitem{schneider2004death}
{Edward~F Schneider}. 2004.
\newblock \showarticletitle{Death with a Story}.
\newblock {\em Human communication research\/} {30}, 3 (2004), 361--375.
\newblock


\bibitem{Shamma:2011:PPM:1958824.1958878}
{David~A. Shamma}, {Lyndon Kennedy}, {and} {Elizabeth~F. Churchill}. 2011.
\newblock \showarticletitle{Peaks and Persistence: Modeling the Shape of
  Microblog Conversations}. In {\em Proc. CSCW}. ACM, New York, NY, USA,
  355--358.
\newblock
\showISBNx{978-1-4503-0556-3}


\bibitem{Sharifi:2010:SMA:1857999.1858099}
{Beaux Sharifi}, {Mark-Anthony Hutton}, {and} {Jugal Kalita}. 2010.
\newblock \showarticletitle{Summarizing Microblogs Automatically}. In {\em
  Proc. HLT}. Association for Computational Linguistics, Stroudsburg, PA, USA,
  685--688.
\newblock
\showISBNx{1-932432-65-5}


\bibitem{wahl2013strategic}
{Karin Wahl-Jorgensen}. 2013.
\newblock \showarticletitle{The strategic ritual of emotionality: A case study
  of Pulitzer Prize-winning articles}.
\newblock {\em Journalism\/} {14}, 1 (2013), 129--145.
\newblock


\bibitem{zhong2013sharing}
{Changtao Zhong}, {Sunil Shah}, {Karthik Sundaravadivelan}, {and} {Nishanth
  Sastry}. 2013.
\newblock \showarticletitle{Sharing the Loves: Understanding the How and Why of
  Online Content Curation}. In {\em Proc. ICWSM} {\em (ICWSM)}.
\newblock


\end{thebibliography}

\end{document}